\documentclass[12pt,twoside]{report}
\usepackage{indentfirst}
\newif\ifpdf
\ifx\pdfoutput\undefined
\pdffalse % we are not running PDFLaTeX
\else
\pdfoutput=1 % we are running PDFLaTeX
\pdftrue \fi
\usepackage[authoryear]{natbib}
\usepackage{esdiff}
\usepackage{ifthen}
\usepackage{amstext}
\usepackage{amssymb}
\usepackage{version}
\usepackage{bm}
\usepackage{type1cm}
\usepackage{euscript}
\usepackage{bbding}
\usepackage{fleqn}
\usepackage{epsfig}
\usepackage{epic,eepic}
\usepackage{float}
\usepackage{color}
\usepackage[unicode,bookmarks,bookmarksopen,bookmarksopenlevel=0,colorlinks,linkcolor=blue,citecolor=red]{hyperref}
\usepackage[fleqn]{amsmath}
\usepackage{mathrsfs}

\AtBeginDocument{%
  \setlength\abovedisplayskip{2pt}
  \setlength\belowdisplayskip{2pt}}

\usepackage{fancyhdr}
\fancypagestyle{plain}{%
\fancyhead{} % get rid of headers
}

\DeclareMathOperator\Div{div}

\DeclareMathOperator{\const}{const}

\newcommand\vu{\bm{u}}

\newcommand\vX{\bm{X}}

\newcommand\mK{\mathbb{K}}
\newcommand\mJ{\mathbb{J}}
\newcommand\mH{\mathbb{H}}
\newcommand\mI{\mathbb{I}}
\newcommand\Tr{\operatorname{Tr}}

\newcommand\vx{\bm{x}}

\newcommand\vxi{\bm{\xi}}

\begin{document}

\DeclareGraphicsExtensions{.jpg,.pdf,.mps,.png}

{ \makeatletter
\renewcommand{\@oddhead}{\hfil\thepage\hfil}
\makeatother \thispagestyle{empty} \vspace*{-10mm}

\begin{center}

{\Large {\bf Continuity equations\\[3mm]
in the Generalised Lagrangian Mean (GLM) theory}}
%:\\[4mm] exact formulations}}
\\[3mm]
{\textsf{by V.\,A.~Vladimirov,\ Universities of York and Leeds}%,\ vv500@york.ac.uk}
\\[3mm]
%{Dated: May-03-2026}
} 

\end{center}
}

%\maketitle
%\tableofcontents

%\addcontentsline{toc}{section}
%\addcontentsline{toc}{section}

%\tableofcontents
%\listoffigures \listoftables

%\addcontentsline{toc}{section}

\section*{Abstract}\label{abstract}
Generalised Lagrangian Mean (GLM) theory describes the joint evolution of the mean flow and its perturbations. 
This paper analyses the exact GLM continuity equations (GLM-CEs), presents them in new forms, and adds new equations to complete them. 
Our results include:
\textbf{(i)} A modified approach to GLM theory that uses only the basic notions of classical fluid dynamics.
The only tools used are Lagrangian $\vX$, Eulerian $\vx$, averaged (mean) Eulerian $\bar\vx$ coordinates of fluid particles, and an averaging parameter $\alpha$.
The targeted forms of GLM-CEs are expressed with functions $\vx(\bar\vx,t,\alpha)$ and $\rho(\bar\vx,t,\alpha)$, where $\rho$ is the fluid density.
\textbf{(ii)} Actual velocity divergence  $\nabla\cdot\vu$ expressed through $\bar\nabla\cdot\bar\vu$, where $\vu$ and $\bar\vu$ are the actual and mean fluid velocities. 
\newline\textbf{(iii)} Three new forms of GLM-CEs. 
Each version represents a sum of a mean and an $\alpha$-dependent (tilde) part. 
The latter addition makes each form mathematically complete.
\textbf{(iv)} The original Andrews-McIntyre transformation (AMT) is also complemented by its tilde part.
Then it also gives the full version of the exact GLM-CEs.
However, it works only for a special class of fluid flows.
Its structure explains why the tilde equations have been overlooked.
\textbf{(v)} Our generalisation of AMT, which repeats the results (i)-(iii) but requires a more complex derivation.
\textbf{(vi)} Examples of flows with small but finite Lagrangian perturbations, linking our exposition to the classical GLM theory.

\emph{Keywords:} continuity equations, general Lagrangian mean theory, average (mean) coordinates, Eulerian description,  tilde equations, Andrews-McIntyre transformation.
\vspace{2mm}

\section{Introduction}

The Generalised Lagrangian Mean (GLM) theory (also known as the hybrid Euler–Lagrange procedure) was developed in [1, 2, 6, 7, 11, 12, 15-17] and other papers and monographs [4, 5].
This theory represents a major attempt in fluid mechanics to derive and use a system of \textit{exact governing equations} that describe the joint evolution of the unsteady mean flow and its Lagrangian perturbations.
Several achievements and problems warrant study.
These include the GLM forms of the continuity equations (CEs), abbreviated here as GLM-CEs.
The three original forms of CEs are as follows
\begin{align}\label{ContEqn-orig}
&& \textbf{(a)} \ \Div \vu=0;\quad \textbf{(b)} \ \Div\vu=0 \ \text{and}\ D\rho=0;\quad \textbf{(c)}\ \Div\vu+ D \ln\rho=0,
\end{align}
where $\vu=\vu(\vx,t)$ and $\rho=\rho(\vx,t)$ are the velocity and density fields, $\vx=(x_1,x_2,x_3)$ and $t$ are the Cartesian coordinates and time, and $D$ is the material derivative.
Eqn.\eqref{ContEqn-orig}\textbf{(a)} is valid for an incompressible fluid with $\rho=\const$, Eqn.\eqref{ContEqn-orig}\textbf{(b)} for an incompressible fluid with $\rho\neq\const$, and Eqn.\eqref{ContEqn-orig}\textbf{(c)} for a compressible fluid, which contains \textbf{(a)} and \textbf{(b)} as special cases.
All Eqs.\eqref{ContEqn-orig} are written in forms that separate the velocity divergence.
No CE constitutes an independent equation; it forms the governing systems of fluid mechanics together with the momentum equation.
Therefore, in this paper, we restrict ourselves only to exact formulae and the most general statements for GLM-CEs.
This paper clarifies the transformations of all CEs \eqref{ContEqn-orig} into their GLM-CE forms and justifies the appearance of their tilde parts.
Our approach uses only the basic notions and methods of classical fluid dynamics [3, 10, 14].
The distribution of material is as follows:

Section 2 introduces the basic notation.
There are three types of coordinates: Lagrangian $\vX$, Eulerian $\vx$, and average Eulerian $\bar\vx$ and an averaging scalar parameter $\alpha$.
The average operation on $\alpha$ is denoted as $\langle\cdot\rangle$.
The keystone of the theory is to move the description of fluid motion from the original independent variables $\vx$ and $\vX$  to the average coordinate $\bar\vx$. 
Any function $F$ in the GLM description can be presented as the sum of its bar and tilde parts:
 \begin{align}\label{F-tilde}
    F(\bar\vx,t,\alpha)=\bar F(\bar\vx,t)+\tilde F(\bar\vx,t,\alpha) \quad \text{where}\quad \bar F =\langle F\rangle \quad\text{and}\quad  \langle \tilde F\rangle=0.
\end{align}
In the following, we call them \textit{bar functions} and \textit{tilde functions}.
In addition, in contrast to the existing GLM literature, we use $\bar\vx$, $\vx$, and $\vX$ flexibly, frequently interchanging them to simplify calculations and interpretations, even within a single formula.

Section 3 contains the main results of the paper.
First, we express the actual velocity divergence \(\nabla\cdot\vu\) in terms of \(\overline\nabla\cdot\overline\vu\) and the Jacobian \(H\) of transformation from \(\vx\) to \(\bar\vx\).
Second, we introduce three forms of exact GLM-CEs corresponding to \eqref{ContEqn-orig}.
Third, we show that each of these equations possesses a special structure of type $\overline\Phi=\widetilde\Psi$ with some functions $\overline\Phi$ and $\widetilde\Psi$,
which entails two equations $\overline\Phi=0$ and $\widetilde\Psi=0$.
The former yields the averaged GLM-CEs, while the latter yields the related tilde equation.
As a result, each exact GLM-CE represents a system of average and tilde parts of the equations.
The latter upgrades the existing GLM governing equation systems by adding the tilde part of GLM-CE to each.
It is important that in the presence of small parameters (e.g. the amplitudes), the orders of the exact, averaged, and tilde equations are the same.
Certainly, the additional tilde GLM-CEs equations vanish after their averaging and do not enter the full average system of GLM equations. 
However, they impose constraints that define the class of admissible functions entering the averaging procedure. 
The same situation occurs for the momentum equation, where the average equations are used together with the original (not averaged) equations for Lagrangian displacements [1, 3-6]. That reflects the well-known fact that the average system of GLM equations is not closed.

Section 4 analyses the Andrews-McIntyre transformation (AMT) of \eqref{ContEqn-orig} into GLM CEs [1, 4, 5, 7], 
We emphasise that the original AMT [1] currently represents the only way to derive GLM CEs.
We give AMT in two versions.
The first represents our generalised form of AMT.
It yields the results of Section 3 but is more complex to derive.
In contrast, the original AMT [1] is based on the introduction of a new auxiliary function $\sigma(\bar\vx,t)$ that represents a purely bar function ($\tilde\sigma\equiv 0$), satisfies a heuristically chosen PDE and special initial conditions $\sigma(\bar\vx,0)=\rho(\bar\vx,0,\alpha)$. 
Hence, it applies to special classes of fluid ensembles.
Its structure explains why [1, 4, 5, 7] occasionally excluded the tilde parts of GLM-CEs.
We add brief conclusions in terms of AMT function $\sigma$.

Section 5 provides examples of flows with small Lagrangian displacements, connecting the paper to classical GLM theory.
We explicitly present the related tilde equations for further use.
The particular examples of solutions (obtained after incorporating the tilde parts of GLM-CEs) to the governing GLM system are not considered, since this requires the use of momentum equations that are beyond the scope of this paper.

\section{Notations}\label{sect2}
\setcounter{equation}{0}

Consider a fluid motion described by two smooth and invertible vector functions,
\begin{eqnarray}\label{Lagr-Eul}
&&\vx=\vx(\vX,t,\alpha),\qquad \bar\vx=\bar\vx(\vX,t),
\end{eqnarray}
where  $\alpha$ is an ensemble parameter.
The first Eqn.\eqref{Lagr-Eul} describes an ensemble of actual fluid motions, represented by the link between the Eulerian $\vx$  and Lagrangian $\vX$ coordinates of the material particles, while the second describes their averaged motion.
These two one-to-one mappings $\vx\Leftrightarrow\vX$ and $\bar\vx\Leftrightarrow\vX$ yield, by excluding $\vX$, the one-to-one correspondence:
\begin{align}\label{strange}
    \vx\Leftrightarrow\bar\vx, \quad\text{or}\quad \ \vx=\vx(\bar\vx,t,\alpha)\ \ \text{and}\ \ \bar\vx=\bar\vx(\vx,t,\alpha),
\end{align}
with an additional scalar parameter $\alpha$.
As a result, any function that describes the motion of a fluid can be expressed through any of the three independent variables $(\vX,t)$, $(\vx,t)$, or $(\bar\vx,t)$; all three are chosen as Cartesian coordinates in related boundless three-dimensional spaces.
Two types related to \eqref{Lagr-Eul},\eqref{strange} velocities are
\begin{equation}
{\vu}\equiv \partial{\vx}/\partial t|_{\vX},\quad\text{and}\quad \bar{\vu}\equiv\partial\bar{\vx}/\partial t|_{\vX},\label{velocities}
\end{equation}
which are \textit{the actual velocity} and \textit{the average velocity}; the subscripts after the bars denote the variables kept constant.
The variables $\bar\vx$ and $\bar\vu$ represent the averages
\begin{eqnarray}\label{assign}
    \bar\vx\equiv\langle\vx\rangle,\quad\text{and}\qquad \bar\vu\equiv\langle\vu\rangle,
\end{eqnarray}
where the average operation $\langle\cdot\rangle$ commutes with all other mathematical operations used.
We also use three different expressions for the same material derivative,
\begin{eqnarray}\label{mat-der}
&& {\partial}/{\partial t}|_{\vX}={\partial}/{\partial t}|_{\vx}+{\vu}\cdot\nabla={\partial}/{\partial t}|_{\bar\vx}+{\bar\vu}\cdot\bar\nabla\ \ \text{or}\ \ D_{\vX}= D=\bar D.
\end{eqnarray}
This symbolic equality of three operators means that changing the independent variables in a related function should automatically involve the corresponding operator forms.

Other notations and agreements are as follows.

$\bullet$  A notation for bar and tilde functions is already introduced in \eqref{F-tilde}.

$\bullet$ The coordinates $\vx$ or $\vX$ will be used actively. 
This means that any formula containing \(\vx\) and \(\vX\) is treated as potentially expressible in \(\bar\vx\) according to \eqref{Lagr-Eul},\eqref{strange},\eqref{mat-der}. 
Similarly to the flexibility in \eqref{mat-der}, we will use symbolic equalities as
\begin{align}\label{symb}
   f(\vx,t,\alpha)=f(\bar\vx,t,\alpha)=f(\vX,t,\alpha).
\end{align}

$\bullet$ We also write \(f^c=f^c(\bar\vx,t,\alpha)\equiv f(\vx(\bar\vx,t,\alpha),t,\alpha)\) for a composite function of $f(\vx,t,\alpha)$ and $\vx(\bar\vx,t.\alpha)$.

$\bullet$  The links between coordinates \(\vx\), \(\bar\vx\) and \(\vX\) are given by \eqref{Lagr-Eul}-\eqref{strange},\eqref{mat-der} and are further characterised by the matrices and their determinants,
\vspace{1mm}
\begin{align}
&\mJ=(J_{ik})\equiv\Bigl(\frac{\partial x_i}{\partial X_k}\Bigr), \qquad \bar\mJ=(\bar J_{ik})\equiv\Bigl(\frac{\partial \bar x_i}{\partial X_k}\Bigr), \qquad 
\mH=(H_{ik})\equiv\Bigl(\frac{\partial x_i}{\partial\bar x_k} \Bigr),  \label{matrices-s}\\
&J\equiv\det\mJ> 0,\qquad \bar J\equiv\det\bar\mJ> 0,\qquad H\equiv\det\mH={J}/{\bar J}> 0,\label{Jacobians-s}
\end{align}

\section{Transformations of continuity equations}
\setcounter{equation}{0}

Our target is to obtain GLM-CEs with unknown functions  $\vx(\bar\vx,t,\alpha)$, $\bar\vu(\bar\vx,t,\alpha)$, and  $\rho(\bar\vx,t,\alpha)$ \eqref{Lagr-Eul} - \eqref{velocities}.

\textbf{3.1. Transformation of velocity divergence.}
Euler's kinematic equalities for Jacobians (see formula (3.8) in [13]) are:
\begin{align}
{(\rm i)}\quad \nabla\cdot\vu = D\ln J, \qquad {(\rm ii)}\quad \bar\nabla\cdot\bar \vu=\bar D\ln\bar J.\label{J-Euler}
\end{align}
We rewrite (i) by adding and subtracting \(\bar D\ln\bar J\), and then use (ii),
\begin{align}\label{eq1}
    \nabla\cdot\vu = (D\ln J-\bar D\ln\bar J)+\bar D\ln\bar J=\bar D\ln H^c +\bar\nabla\cdot\bar\vu.
\end{align}
This gives the required link between the two divergence forms and the Jacobian $H$.
As a result, GLM-CEs follow after substituting $\nabla\cdot\vu$ from \eqref{eq1} into each equation \eqref{ContEqn-orig}.
In particular, for \eqref{ContEqn-orig}\textbf{(c)}, it gives 
\begin{align}
&D \ln \rho=-\nabla\cdot\vu=-\bar D\ln H^c -\bar\nabla\cdot\bar\vu,\qquad \text{or}\label{CE-transf-0}\\
&\overline\nabla \cdot \overline\vu=Q;\qquad Q\equiv -\bar D\ln(\rho H)^c \label{CE-transf-1},
\end{align}
where \eqref{CE-transf-1} represents the exact GLM-CE\textbf{(c)} required.

\vspace{1mm}
\textbf{3.2. The List of GLM-CEs} is given below using the bar and tilde notation \eqref{F-tilde} and the shortcuts
\begin{align}\label{L-R}
    h\equiv\ln H,\qquad \lambda\equiv \ln\sigma,\qquad \sigma\equiv \rho H.
\end{align}

\textbf{(a)} For \textit{a homogeneous incompressible fluid} \eqref{ContEqn-orig}\textbf{(a)}, it is
\begin{align}
    &\bar\nabla\cdot\bar\vu=-\bar D\, h^c,\label{CE-a}\\
    &\bar\nabla\cdot\bar\vu= -\bar D\, \overline{h^c}, \label{CE-a-av}\\
    &\bar D\, \widetilde{h^c}=0, \label{CE-a-diff}
\end{align}
where the first line gives the exact GLM-CE, the second is its average, and the third is the difference between the previous two, written using the tilde notation \eqref{F-tilde}.

\textbf{(b)} \textit{For an inhomogeneous incompressible fluid} \eqref{ContEqn-orig}\textbf{(b)}, GLM-CE is given by a system of two equations (where to obtain the equation for density, we use \eqref{mat-der}),
\begin{align}
   &\bar D \rho^c=0 \ \ \text{and}\ \ \bar\nabla\cdot\bar\vu= -\bar D\, {h}^c,\label{CE-b}\\
  &\bar D \overline{\rho^c}=0\ \text{and}\ \ \bar\nabla\cdot\bar\vu =-\bar D\,\overline{h^c},\label{CE-b-av}\\
  &\bar D \widetilde{\rho^c}=0 \ \ \text{and}\  \ \bar D\,\widetilde{h^c}=0. \label{CE-b-diff}
\end{align}

\textbf{(c)} \textit{For a compressible fluid} \eqref{ContEqn-orig}\textbf{(c)}, is
\begin{align}
   &\bar\nabla\cdot\bar\vu = - \bar D\, \lambda^c,\label{CE-c}\\ 
   &\bar\nabla\cdot\bar\vu = - \bar D\, \overline{\lambda^c},\label{CE-c-av}\\
   &\bar D\,\widetilde{\lambda^c}=0.\label{CE-c-diff}
\end{align}
The meaning of different lines in \textbf{(b)} and \textbf{(c)} is the same as in \textbf{(a)}.

\vspace{3mm}
\textbf{3.3. The role of tilde parts of GLM-CEs.}
%The first aim of GLM theory is to obtain the system of exact governing equations based on independent variables $(\bar\vx,t,\alpha)$.
The exact GLM-CEs are given by \eqref{CE-a},\eqref{CE-b},\eqref{CE-c}. 
Each of these equations represents a system of two equations: an average (bar) equation (taken as one of \eqref{CE-a-av},\eqref{CE-b-av},\eqref{CE-c-av}) and the related tilde equation taken from \eqref{CE-a-diff},\eqref{CE-b-diff},\eqref{CE-c-diff}.
So far, only the bar parts of the equations have been used in the governing GLM system; see [1, 4, 5, 7].
%They have been derived by the original MAT procedure (see subsection 4.2 below), for the special case of small Lagrangian displacements.
%which was built to avoid the tilde equations.
%derived the averaged equations directly, without concern for the tilde equations. 
%As a result, the tilde equations were not taken into account.
In \eqref{CE-a}-\eqref{CE-c-diff} the role of the tilde equations is restored; they define the classes of admissible fluid motions satisfying the equations
\begin{align}
    &\bar D\, \widetilde{h^c}(\bar\vx,t,\alpha)\equiv\bar D\,\widetilde{\ln H^c}=0\qquad \text{for the cases}\ \textbf{(a)}\ \text{and}\ \textbf{(b)},\label{comp-Eqn}\\
    &\quad \bar D\,\widetilde{\lambda^c}(\bar\vx,t,\alpha)\equiv\bar D\,\widetilde{\ln(\rho H)^c}=0\qquad \text{for the case} \ \textbf{(c)},\label{comp-Eqn2}
\end{align}
where, generally $H=\bar H+\tilde H$ and $\rho=\bar\rho+\tilde\rho$.
Due to the definitions \eqref{Lagr-Eul},\eqref{strange}, the functions $\rho(\bar\vx,t,\alpha)$ and $\vx(\bar\vx,t,\alpha)$ (which form the Jacobian $H$ \eqref{matrices-s},\eqref{Jacobians-s}) essentially depend on $\alpha$; therefore, these equations represent mathematically valuable constraints (see section 5 for examples).
A particular case of \eqref{comp-Eqn},\eqref{comp-Eqn2} is the following,
\begin{align}\label{special1}
     H=\widetilde H(\alpha),\qquad  \rho H=(\widetilde{\rho H})(\alpha)
\end{align}
where the arbitrary functions of $\alpha$ assign different constants to each flow in the ensemble.
An even more specific example is the following. 
\begin{align}\label{H-const}
H=\widetilde{H}\equiv 0,\ \text{for}\ \textbf{(a)}\ \text{and}\ \textbf{(b)}\quad \text{and}\quad \rho H=\widetilde{\rho H}\equiv 0,\ \text{for}\ \textbf{(c)}.
\end{align}
%which makes \eqref{comp-Eqn},\eqref{comp-Eqn2} identically satisfied. 
Both \eqref{special1} and \eqref{H-const} are applicable to specific classes of fluid motion, while Eqs.\eqref{CE-a-av},\eqref{CE-b-av},\eqref{CE-c-av},\eqref{comp-Eqn},\eqref{comp-Eqn2} represent the exact GLM-CEs\textbf{(a)-(c) } that hold for any flow.

%\vspace{-5mm}
The dynamic meaning of the constraints \eqref{comp-Eqn},\eqref{comp-Eqn2} can be understood by rewriting them in Lagrangian coordinates, say $\widetilde{h}(\vX,t,\alpha)$ and integrating  over time,
\begin{align}\label{constraint-Lagr}
    {\partial\, \widetilde h}/{\partial t}|_{\vX}=0,\qquad \widetilde h=\widetilde\chi(\vX,\alpha),
\end{align}
with an arbitrary function \(\widetilde\chi(\vX,\alpha)\).
This demonstrates the conservation of \(\widetilde h\) (and \(\widetilde \lambda\)) in each material particle in each individual flow of an ensemble.
This means that \(\widetilde h\) and \(\widetilde \lambda\) can serve as Lagrangian coordinates.
Certainly, all the tilde GLM-CEs derived above vanish after their averaging and do not enter the average system of GLM governing equations. However, they impose constraints that define the class of functions admitted to the averaging procedure. 
 The same situation occurs for the momentum equation, where the average equations are used together with the original (not averaged) equations for Lagrangian displacements [1, 4, 5, 7]. 
 This reflects that the averaged system of governing GLM equations is not closed.

%\newpage
\section{Andrews-McIntyre Transformation (AMT)}
\setcounter{equation}{0}

\textbf{4.1. The generalised version of AMT.}
The required general equations are as follows. 
\begin{eqnarray}
&& \textbf{({i})} \, D\rho+ \rho\,\nabla\cdot\vu=0;\ \textbf{({ii})} \, DJ=J\,\nabla\cdot\vu; \ \textbf{({iii})} \,D(\rho J)=\partial (\rho J)/\partial t|_{\vX}=0,\label{rhoJ}\\
&& \textbf{({i})} \, \bar D\sigma+\sigma\,\bar\nabla\cdot\bar\vu=0;\ \textbf{({ii})} \, \bar D\bar J=\bar J\,\bar\nabla\cdot\bar \vu;\ \textbf{({iii})} \,  \bar D(\sigma \bar J)=\partial (\sigma\bar J)/\partial t|_{\vX}=0,\label{rhoJbar}
\end{eqnarray}
where $\vx$, $\bar\vx$, and $\vX$ are the coordinates \eqref{Lagr-Eul}, with Jacobians \eqref{Jacobians-s}.
Eqs. \textbf{(i)},\textbf{(ii)},\textbf{(iii)} are the following.

$\bullet$ Eqn.\eqref{rhoJ}\textbf{(i) }is the same as \eqref{ContEqn-orig}\textbf{(c)}.

$\bullet$ Eqn.\eqref{rhoJbar}\textbf{(i)} plays a central role in AMT [1, 4, 5, 7].
It is designed and postulated heuristically, and a new unknown auxiliary function $\sigma$ replaces the density $\rho$ in the GLM governing equations.
According to the theory of Sect.3, it is identical to Eqn.\eqref{CE-c} with $\lambda\equiv\ln\sigma$, which clarifies the meaning of Eqn.\eqref{rhoJbar}\textbf{(i)} and the field $\sigma$ itself.

$\bullet$ Both Eqs.\textbf{(ii)} are the same as \eqref{J-Euler}.

$\bullet$ Both Eqs.\textbf{(iii)} follow automatically from the related Eqs.\textbf{(i)} and \textbf{(ii)}.

\noindent
Both Eqs.\textbf{(iii)} can then be integrated in Lagrangian coordinates as
\begin{align}
\rho(\vX,t,\alpha) J(\vX,t,\alpha) =\rho_0(\vX,\alpha),\qquad \sigma(\vX,t,\alpha) \bar J(\vX,t)=\sigma_0(\vX,\alpha)\label{integrals}
\end{align}
with arbitrary initial data
\begin{align}
&\rho(\vX,0,\alpha)\equiv\rho_0(\vX,\alpha)>0, \qquad \sigma(\vX,0,\alpha)\equiv\sigma_0(\vX,\alpha)>0,\label{initial-data}
\end{align}
Other initial data used are
\begin{eqnarray}\label{JJ=1}
    \vx(\vX,0,\alpha)=\bar\vx(\vX,0) =\vX, \quad J(\vX,0,\alpha)=\bar J(\vX,0)=H(\vX,0,\alpha)=1.
\end{eqnarray}
The ratio of the two equations \eqref{integrals} gives the equality,
\begin{align}\label{continuity-compressible}
\frac{\rho(\vX,t,\alpha)}{\rho_0(\vX,\alpha)}\, H(\vX,t,\alpha) =\frac{\sigma(\vX,t,\alpha)}{\sigma_0(\vX,\alpha)},
 \end{align}
which contains arbitrary functions $\rho_0$ and  $\sigma_0$.
The AMT choice is 
\begin{align}\label{restriction-rho}
    \sigma(\vX,0,\alpha)=\rho(\vX,0,\alpha)\quad \text{or}\quad \sigma_0(\bar\vx,\alpha)=\rho_0(\bar\vx,\alpha).
\end{align}
which transforms \eqref{continuity-compressible} to
\begin{eqnarray}\label{sigma-rho-5}
&& \sigma^c=(\rho H)^c\quad\text{or}\quad \sigma=\rho H,
 \end{eqnarray}
that is already derived by a different method in \eqref{L-R}.
Hence, the generalised AMT form of Eqn.\eqref{ContEqn-orig}\textbf{(c)} consists of two equations,
\begin{align}\label{continuity-compressible2}
&\rho=\sigma/H,\qquad\bar D\sigma+\sigma{\bar\nabla}\cdot{\bar\vu}=0,\qquad \text{which can be rewritten as}\\
&\bar\nabla\cdot\bar\vu=Q,\quad Q\equiv-\bar D \ln\sigma^c=- \bar D \ln (\rho H)^c,\label{rhoJbar1}
\end{align}
returning us to $Q$ \eqref{CE-transf-1}. 
Hence, the generalised ATM produces the same results as in Section 3 (including the necessity of the bar and tilde equations \eqref{comp-Eqn},\eqref{comp-Eqn2}), but requires additional details and effort in the derivation procedure \eqref{rhoJ}-\eqref{rhoJbar1}. 

\vspace{1mm}
\textbf{4.2. The original version of AMT}, see  [1, 3-6], represents a special case of \eqref{rhoJ}-\eqref{rhoJbar1} that satisfies additionally enforced restrictions,
\begin{align}
     &\sigma=\sigma^c\equiv\bar\sigma(\bar\vx,t),\quad \widetilde\sigma(\bar\vx,t,\alpha)\equiv 0;\label{sigma-MAT}\\
     &\bar\sigma(\bar\vx,0)=\bar\rho(\bar\vx,0),\quad \widetilde\rho(\bar\vx,0,\alpha)\equiv 0.\label{BC-MAT}
\end{align} 
where \eqref{sigma-MAT} means that $\sigma$ by definition represents a purely bar function, while \eqref{BC-MAT}
introduces an additional constraint when all initial density distributions in the ensemble are the same.
Therefore, by definition \eqref{sigma-MAT}, the source term $Q$ in \eqref{rhoJbar1} also represents a mean value, eliminating the need for tilde equations of type \eqref{comp-Eqn}.
However, different tilde equations automatically appear as the tilde part of \eqref{sigma-rho-5} using \eqref{sigma-MAT}, \eqref{BC-MAT}.
It yields
\begin{align}
 \widetilde H=0,\ \text{for cases}\ \textbf{(a)}\ \text{and}\ \textbf{(b)},\qquad  \widetilde {\rho H}=0,\ \text{for case }\ \textbf{(c)}, \label{sigma-physics}
\end{align}
which coincide with \eqref{H-const}.
These equations, together with \eqref{continuity-compressible2} \eqref{rhoJbar1},
form a system of exact GLM-CEs using the original AMT. 
These tilde equations were not considered in [1, 3-6].
This is because in [1], the GLM-CEs were built as a purely averaged equation using the function $\sigma\equiv\bar\sigma$, postulated to not have the tilde part.
The automatic appearance of the tilde part in \eqref{sigma-rho-5} in this case was overlooked.

The conclusions of this section can be formulated in terms of the AMT function $\sigma$.
The implications of the above results may appear in any part of the general GLM theory where the density $\rho$ or the auxiliary function $\sigma$ is involved.
The two options are the following.
Choosing $\sigma\equiv\bar\sigma$ as a purely bar function \eqref{sigma-MAT} (following [1, 4, 5, 7]) leads to two outcomes: (i) the class of fluid ensembles under consideration is constrained by  \eqref{BC-MAT}, and (ii) an equation \eqref{sigma-physics} should be added to the governing system of GLM equations.
 On the other hand, choosing the full scale function $\sigma=\bar\sigma+\tilde\sigma$ allows one to consider any fluid motions but leads to the appearance of new terms in averaged GLM governing equations.
 It may affect both the momentum equation and the principle of wave action [2, 4, 5].
 This case has no advantage in introducing $\sigma$ compared to $\rho$.

\vspace{-5mm}
\section{Examples of small Lagrangian perturbations}
\setcounter{equation}{0}

The classical GLM theory [1] is based on presenting the functions
\begin{align}\label{strange+}
     \vx=\vx(\bar\vx,t,\alpha)\  \ \bar\vx=\bar\vx(\vx,t,\alpha), \ \ \text{and}\ \vxi\equiv \vx-\bar\vx.
\end{align}\
To establish a link between the above results and the classical GLM theory [1, 4, 5, 7], it is worth decomposing \(Q\equiv-\bar D\ln(\rho H)^c\) \eqref{CE-transf-1} and $H$ \eqref{sigma-physics} for small Lagrangian displacements \(\vxi\), which is understood as the smallness of the matrix \( \left \Vert \mK \right \Vert =O(\varepsilon)\) or its all components $|\xi_{i,k}|=O(\varepsilon)$, where
\begin{align}
    \mK\equiv\mH-\mI=(\partial\xi_i/\partial \bar{x}_k)\equiv (\xi_{i,k}),\quad \mI\equiv(\delta_{ik}).
\end{align}
The decomposition of  $Q$  (for cases \textbf{(a)} and \textbf{(b)}) as a series in $\bigl(\operatorname{Tr}\mK^n\bigr)$ is
  \begin{align}\label{series-main-1}
 Q=-\bar D \ln H^c=\bar D  {\sum_{n=1}^\infty}{(-1)^{n}\Tr(\mK^n)/n}; \qquad  n=1,2,3,\dots,
  \end{align}
where \(\Tr\) marks the trace of a matrix; for the derivation, see [7, 8, 12].
In quadratic precision, it gives the following:
\begin{align}
& Q=\bar D [\operatorname{Tr}(-\mK+\tfrac12\mK^2) +O(\varepsilon^3)\big]=\bar D (-\tilde\xi_{i,i}+\tfrac12 \tilde\xi_{i,k}\tilde\xi_{k,i})+O(\varepsilon^3),\label{Q-quad}\\
&\langle Q\rangle=\tfrac12\bar D\langle \tilde\xi_{i,k}\tilde\xi_{k,i}\rangle+O(\varepsilon^3),\label{Q-quad-av}
\end{align}
where the summation convention is used.
We follow the basic GLM agreement by replacing $\vxi$ with $\tilde\vxi$, such that $\langle\tilde\vxi\rangle\equiv 0$ and $\bar\vxi\equiv 0$.
Then the approximation for the compatibility equations \eqref{comp-Eqn} is
\begin{align}
  \tilde Q\equiv Q-\langle Q\rangle= \bar D\bigl( -\tilde\xi_{i,i}+\tfrac12 \widetilde{\tilde\xi_{i,k}\tilde\xi_{k,i}}\bigr)+O(\varepsilon^3)= 0\quad \text{where}\quad \tilde\vxi=\tilde\vxi(\bar\vx,t,\alpha).
\end{align}
The tilde equation \eqref{sigma-physics} for the Andrews-McIntyre class of flows  can be written as an infinite series for $\tilde H=0$ where
\begin{align}
H\equiv\det(\mI+\mK)=\exp\!\left(\sum_{n=1}^{\infty}(-1)^{n+1}\Tr(\mK^n)/n\right),\label{H-inf}
\end{align}
or its polynomial version, 
\begin{align}
 {H}= 1 + \Tr(\mK) + \tfrac12 [(\Tr\mK)^2 - \Tr(\mK^2)] + O(\varepsilon^3).\label{H-4}
\end{align}
%\quad  \frac16\Bigl((\Tr\mK)^3- 3\,\Tr(\mK)\Tr(\mK^2)+ 2\,\Tr(\mK^3)\Bigr)
%For the method of successive approximations, the different order terms in $\tilde Q=0$ and $\tilde H=0$ vanish independently for each $n$.
More complex equations (involving the density field) for \eqref{comp-Eqn2},\eqref{H-const},\eqref{sigma-physics} follow automatically with $H=\bar H+\tilde H$ and $\rho=\bar\rho+\tilde\rho$.
They are more bulky and are not given here.

%\vspace{1mm}
The author thanks Professor O. Bühler, Professor A. D. D. Craik, and Professor H. K. Moffatt for many useful discussions.

\vskip 5mm
%\newpage

{\textbf{References:}}
\vskip 1mm
%\bibitem{McIntyre}  
1. Andrews, D.G and McIntyre, M.E. 1978. An exact theory of non-linear waves on a Lagrangian-mean flow.
{\it J. Fluid Mech.}, \textbf{89}, 609-646.
\vskip 1mm

2. Andrews, D.G. and McIntyre, M.E. 1978. On wave-action and its relatives. \textit{J. Fluid Mech.} \textbf{89}(4), 647-664.
\vskip 1mm

3. Batchelor, G.K. 2000. An introduction to fluid dynamics. CUP.
\vskip 1mm

4. Bühler, O. 2009. \textit{Waves and mean flows}, CUP.
\vskip 1mm

5. Craik, A.D.D. 1985. \emph{Wave interactions and fluid flows}, CUP.
\vskip 1mm

6. Gilbert, A.D. and Vanneste, J. 2025. Geometric approaches to Lagrangian averaging. \textit{Annual Review of Fluid Mechanics}, \textbf{57}, 117-140.
\vskip 1mm

%\bibitem{Grimshaw} 
7. Grimshaw, R. 1979. Mean flows induced by internal gravity wave packets propagating in a shear flow. \textit{Phil. Trans.  Royal Society of London A}, \textbf{292}(1393), 391-417.
\vskip 1mm

8. Johnson, C.R. and Horn, R.A., 1985. \textit{Matrix analysis.} CUP.
\vskip 1mm

9. Infinite series \eqref{series-main-1} was obtained analytically and verified by artificial intelligence.

10. Lamb, H. 1879. A Treatise on the Mathematical Theory of the Motion of Fluids. The University Press.
\vskip 1mm

11. Maitland-Davies, C. and Bühler, O. 2023. Two-way wave–vortex interactions in a Lagrangian-mean shallow water model. \textit{Journal of Fluid Mechanics}, \textbf{954}, p.A1.
\vskip 1mm

12. Moffatt, K. and Dormy, E. 2019. \textit{Self-exciting fluid dynamos}, CUP.
\vskip 1mm

13. Petersen, K.B. and Pedersen, M.S., 2008. \textit{The matrix cookbook.} Technical University of Denmark, \textbf{7}(15).
\vskip 1mm

14. Serrin, J., 1959. Mathematical principles of classical fluid mechanics. \textit{In Fluid Dynamics I/Strömungsmechanik I}, 125-263, Springer.
\vskip 1mm

15. Soward, A.M., 1972. A kinematic theory of large magnetic Reynolds number dynamos. \textit{Phil. Trans. Royal Society of London A}, \textbf{272}(1227), 431-462.
\vskip 1mm

16. Soward, A.M. and Roberts, P.H. 2010. The hybrid Euler–Lagrange procedure using an extension of Moffatt's method. \textit{Journal of Fluid Mechanics}, \textbf{661}, 45-72.
\vskip 1mm

17. Vladimirov, V. A. 2026. Discover the GLM and pseudo-Lagrangian equations of fluid dynamics on four pages. \textit{ArXiv}, https://arxiv.org/abs/2601.13237.

%\end{thebibliography}

\end{document}